\documentclass[11pt]{article}
\usepackage{moriond,epsfig}

\def\Journal#1#2#3#4{{#1} {\bf #2}, #3 (#4)}

\def\PLB{{\em Phys. Lett.}  B}

\def\be{\begin{equation}}
\def\ee{\end{equation}}
\def\bea{\begin{eqnarray}}
\def\eea{\end{eqnarray}}

\def\ppbar{p\bar{p}}

\begin{document}
\vspace*{4cm}
\title{HIGGS SEARCHES AT THE TEVATRON}

\author{ QIZHONG LI \\ 
(For the CDF and D\O\ Collaborations)}

\address{Fermi National Accelerator Laboratory, P.O. Box 500, Batavia, IL 60510, USA}

\maketitle\abstracts{
One of the highest priority physics goals for the upgraded Tevatron
experiments, CDF and D\O, is the search for the Higgs boson.
We present the initial results from both experiments, based on 40--90 pb$^{-1}$
integrated luminosity, of Higgs searches in several final states, 
including $WH$ and $ZH$, $H\to WW$, and doubly-charged Higgs.
}

\section{Introduction}

A major goal of Tevatron Run II is the search for the Higgs boson. Direct
searches at LEP have excluded a Standard Model Higgs having a mass below
114.4 GeV at 95\% confidence level, with a hint of an excess just above
that. Indirect evidence from a global Standard Model fit to
electroweak data can also be used to constrain the Higgs boson mass.
The LEP Electroweak Working Group recently reported,
that the global fit to LEP electroweak data gives a Higgs boson mass
of $91^{+58}_{-37}$ GeV with an upper limit of 211 GeV at 95\% C.L.
\cite{lathuile}.  These results suggest that the Higgs boson mass may
not be very high and add urgency to Higgs boson searches at the
Tevatron for Run II.

The Higgs boson production mechanism with the largest cross section,
$\sim 0.7$pb for a Higgs mass of 120 GeV, is gluon fusion~\cite{SHWG}.
Unfortunately the background in this mode is very large.  
The most promising Standard Model Higgs discovery
channels at the Tevatron is through associated production with
$W/Z$, where $W/Z$ decays leptonically. The Higgs boson production
cross-section is $\sim0.16$ pb for $WH$ and $\sim0.1$ pb for $ZH$, 
for a Higgs boson mass of 120 GeV.
It is worth to mention that although the $ZH$ cross section is roughly a
factor of two lower over the same Higgs mass range than the $WH$
production cross section, the $Z$ decays to neutrino pairs give a larger
cross section times branching ratio than a single lepton $W$ decays.
The QCD corrections to $\sigma (q\bar q \to WH { \rm  or } ZH)$ coincide with
those of the Drell-Yan process and increase the cross section by
about 30\% \cite{ref1}$^{,}$ \cite{ref2}$^{,}$ \cite{ref3}.  In order to discover a Higgs signal in
this channel at the Tevatron, the main challenge is to be able to
separate the signal from an irreducible Standard Model $Wb\bar b$ or
$Zb\bar b$ background.

\begin{figure}
\centerline{
\psfig{figure=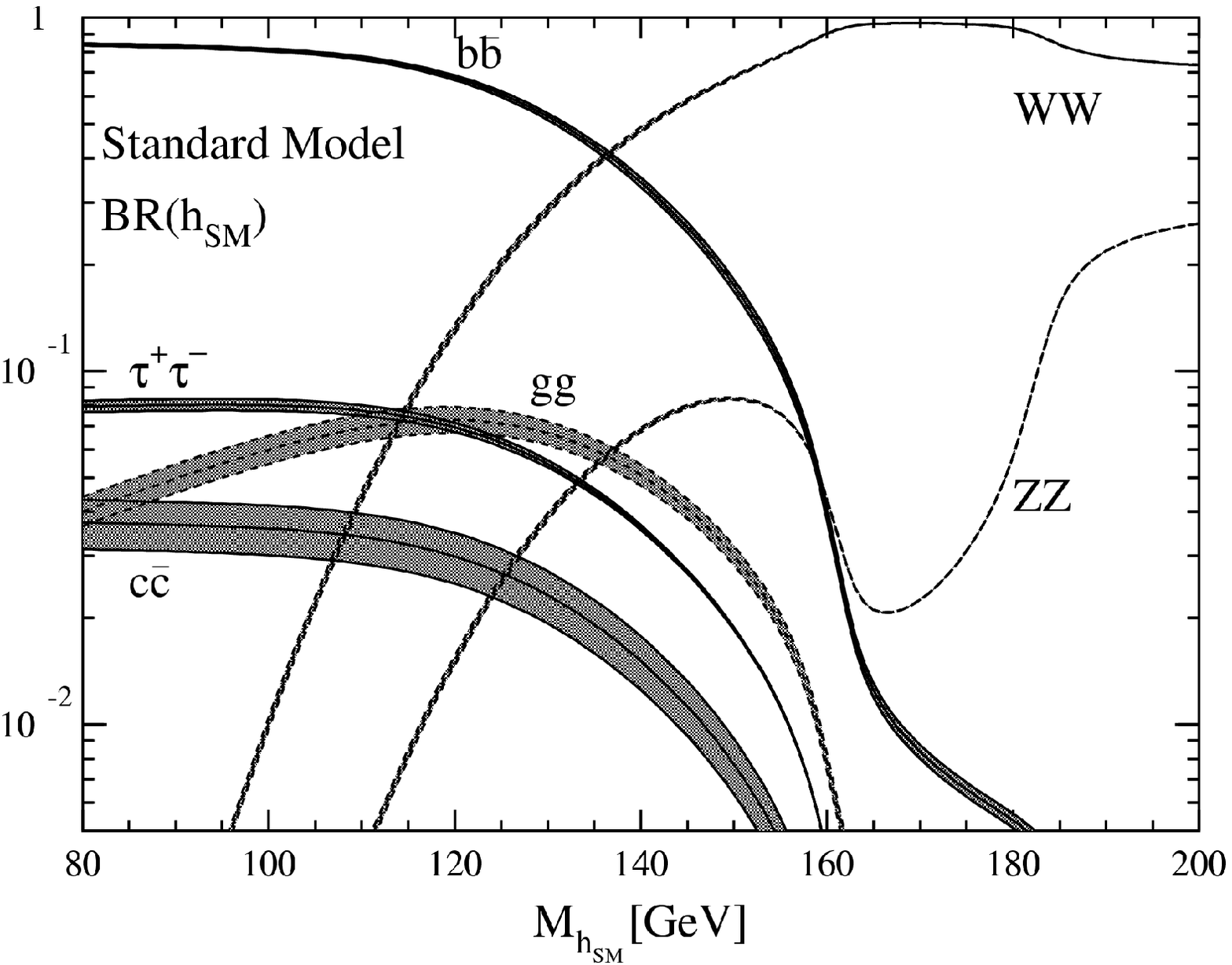,height=1.5in}
\hskip 3.0cm
\psfig{figure=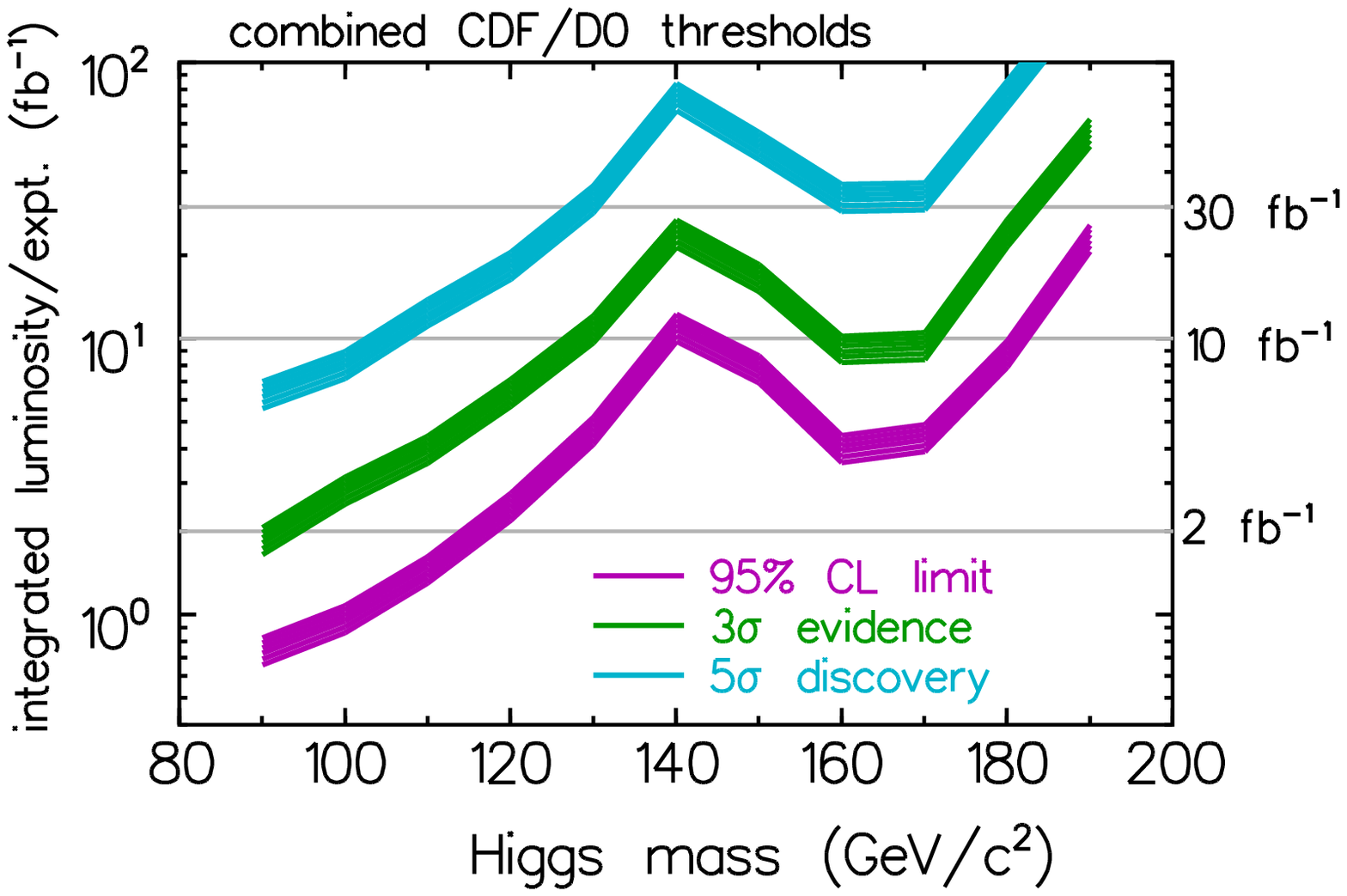,height=1.5in}}
\vskip -0.1cm
\centerline{a) {\hskip 9cm} b)}
\vskip -0.1cm
\caption{a) Branching ratios of the dominant decay modes of the SM Higgs boson,
b) the integrated luminosity required per experiment, to either exclude a 
SM higgs at 95\% C.L. or discover it at $3\sigma$ or $5\sigma$ level.
\label{fig:higgs}}
\end{figure}

The branching ratios for the dominant decay modes of a Standard Model
Higgs boson are shown as a function of Higgs boson mass in
Fig.~\ref{fig:higgs} a). For Higgs boson masses below about 135 GeV, the
decay of $H\to b\bar b$ is dominant. For Higgs masses above 135 GeV,
$H\to WW^*$ becomes the dominant mode.


Using Run I data, CDF has searched for $WH$ and $ZH$ in different
channels, including $Z$ decaying to dileptons, $W$ decaying to leptons
or hadrons, with Higgs decay to $b\bar b$. The result from combining
these channels gives 
$$ \sigma(p\bar p\to VH) * Br(H\to b\bar b) < 7.4 \hbox{pb \quad \quad 
at 95\% C.L.}$$

A few years ago, a Tevatron Higgs Working Group's study evaluated
the Higgs discovery potential for the Tevatron Run II. This was a
joint effort of theorists and experimentalists from both CDF and 
D\O\ experiments. The study was based on a parameterized detector
simulation. The main conclusions
\cite{SHWG} are  that to maximize the Higgs discovery at the
Tevatron, one must combine data from
both experiments, CDF and D\O; must combine all channels, and must
improve the understanding of signal and background processes as well as 
improve the detector performance. Figure.~\ref{fig:higgs} b) shows the
integrated luminosity per experiment for a 95\% C.L. exclusion of a SM
Higgs boson or a $3\sigma$ or $5\sigma$ discovery.  Recently a new
group, the Higgs Sensitivity Group, has been formed which is
reevaluating the Tevatron Run II sensitivity 
using more realistic simulations.

\section{Tevatron and Detector upgrades}

The Tevatron in Run II is the world's highest energy accelarator.  With
the Tevatron upgrades, a new Main Injector,
this machine can deliver to the experiments an order of
magnitude more instantaneous luminosity, as compared with Run I. 
This greatly
enhances the discovery potential for the Higgs boson, since the
discovery reach is limited by the integrated luminosity.  In Run I the
CDF and D\O\ experiments each collected data corresponding to about 
0.1 fb$^{-1}$  of integrated
luminosity. In Run II we expect to collect 5 fb$^{-1}$, 
with the possibility of 10 fb$^{-1}$ if we run efficiently 
through the end of the decade. So far, each
experiment has recorded about 100 pb$^{-1}$ data, with a recent peak
luminosity of $4.0\times 10^{31} /{\rm cm}^2/{\rm sec}$.
In Run II the machine energy is increased from 1.8 TeV to 1.96 TeV in
the $p\bar p$ center of mass; this typically increases physics cross 
sections by about 30--40\%.

Both CDF and D\O\ have upgraded their detectors. CDF has a new central
drift chamber and silicon tracker, new forward calorimeters covering
the psuedorapidity range from 1 to 3, new time of flight detector,
extended muon coverage, and an improved tracking and secondnary
vertex trigger. D\O\ has a new silicon and fiber tracker in 2 Telsa
magnetic field, added preshower detectors in front of the calorimeter,
a much improved muon system, and a new data aquisition and trigger
systems. The upgrades of the detectors enhance the ability to make
sensitive searches.

\section{Run II Analyses and Results}

\subsection{$WH$ and $ZH$ analyses}
\begin{figure} 
\centerline{
\psfig{figure=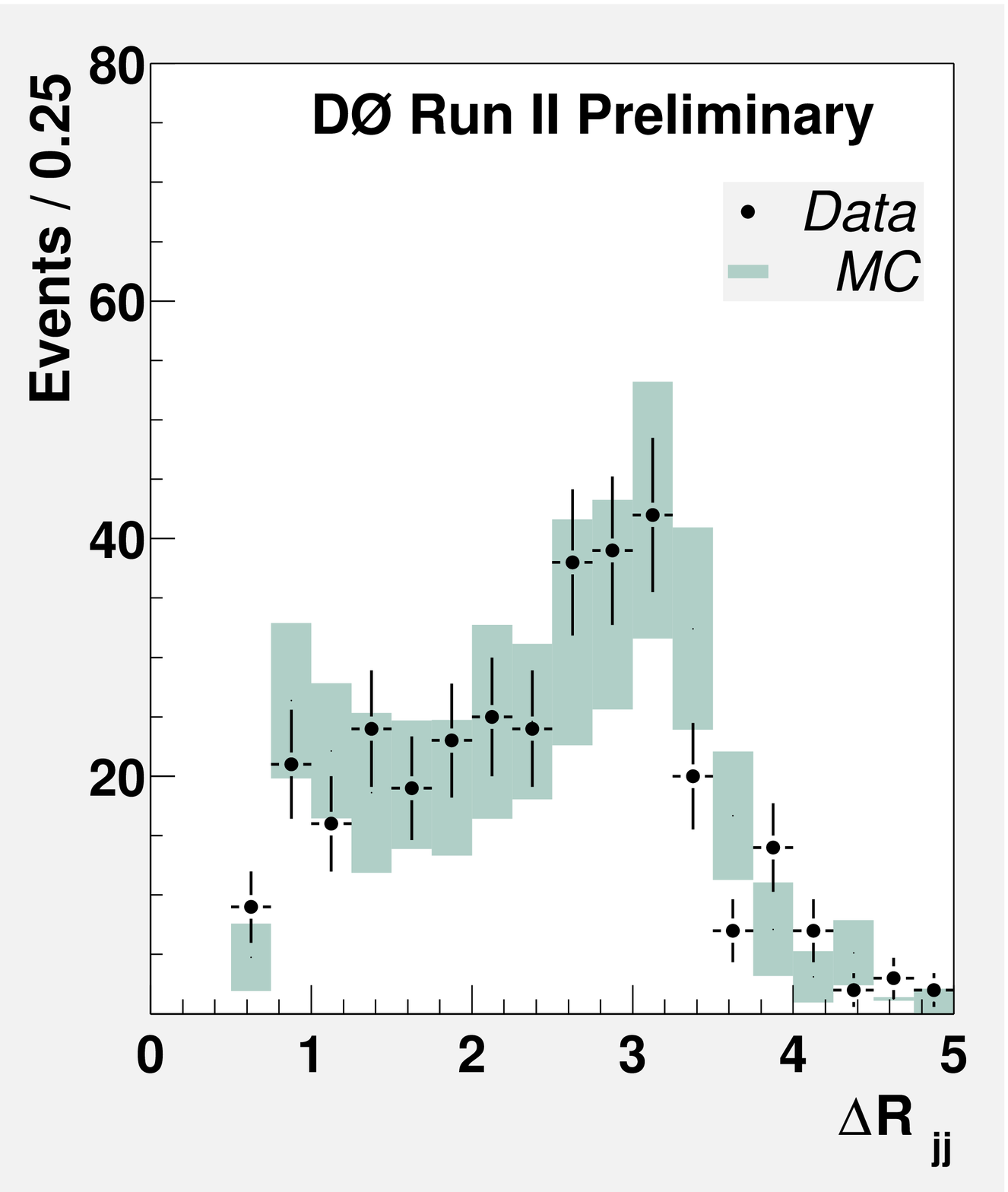,height=2.0in}
\psfig{figure=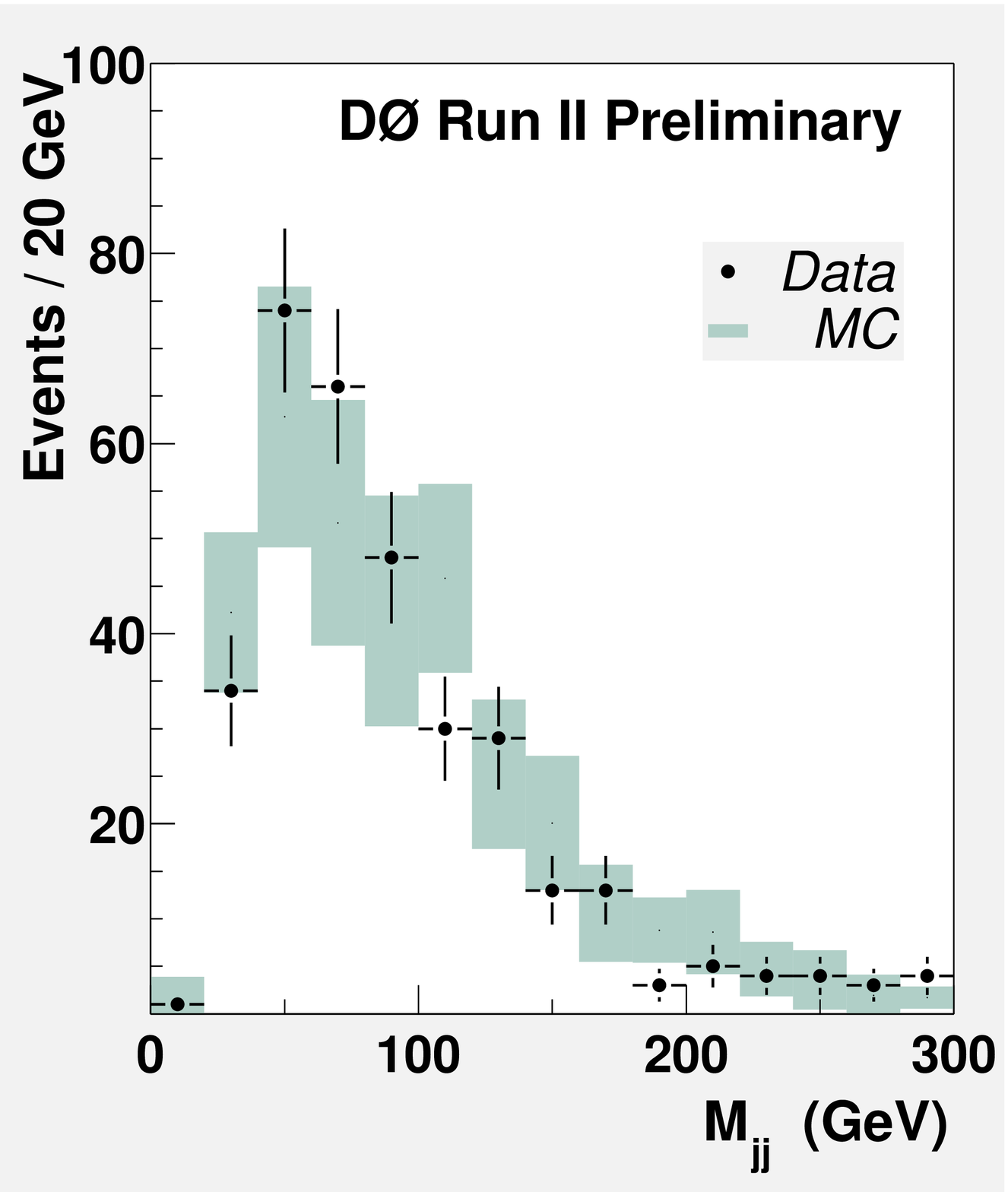,height=2.0in}
\psfig{figure=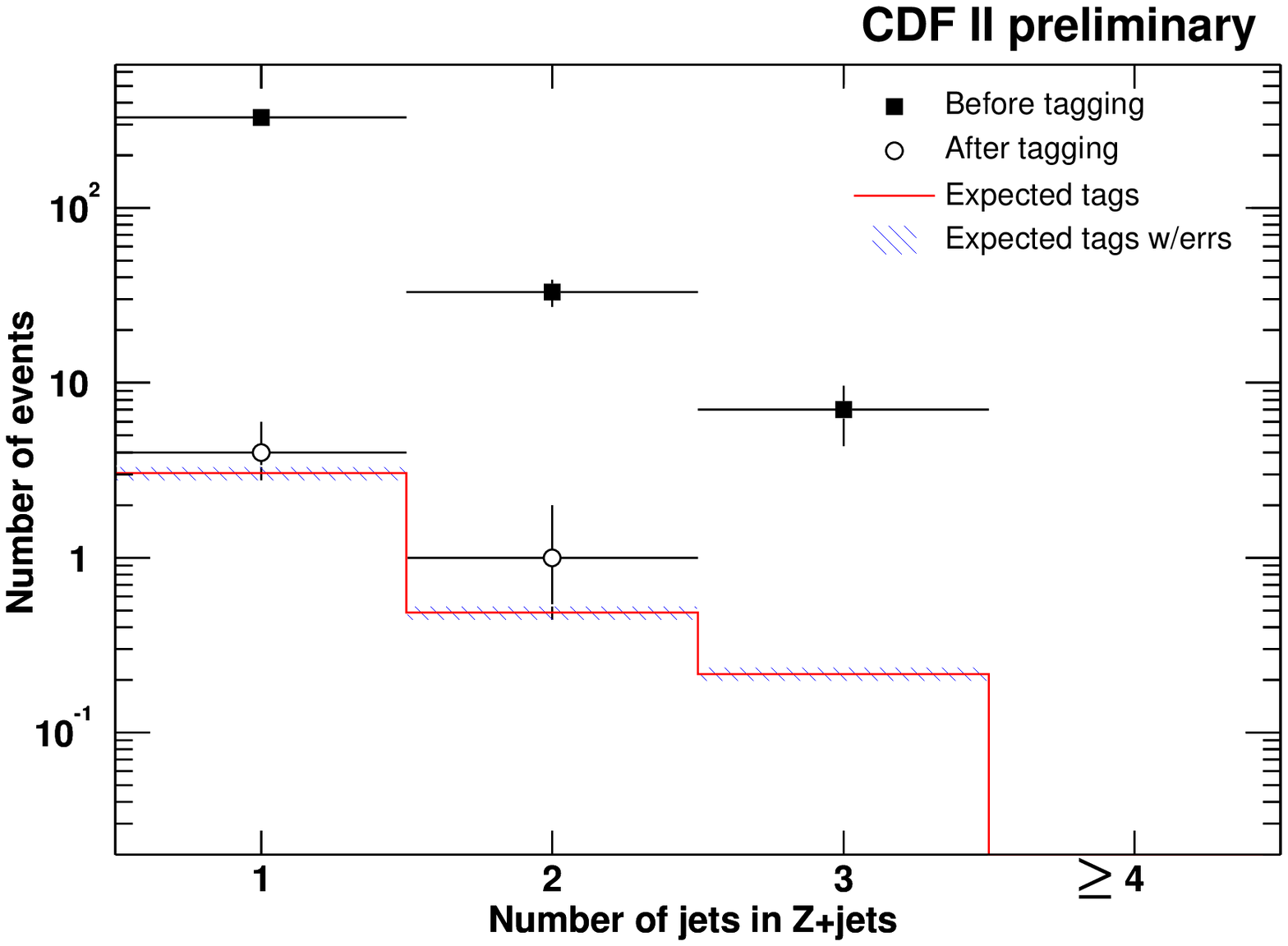,height=2.0in}}
\vskip -0.1cm
\centerline{a) {\hskip 5cm} b) {\hskip 5cm} c)}
\vskip -0.1cm
\caption{a) and b) are $\Delta R_{jj}$ and $M_{jj}$ distributions for 
D\O\ $W$+jets sample. The dots are data and the shaded bands are Pythia MC, 
including systematic errors. 
c) is jet multiplicity distribution for CDF Z+jets sample, before and after
b-tagging.
\label{fig:wz}}
\end{figure}

The first step towards the 
$WH \to \ell\nu + bb$ and $ZH \to \ell\ell + b\bar b$ 
searches is to study the $W+{\rm jets}$ and $Z+{\rm jets}$ samples,
which are the major backgrounds for these analyses.
The basic event selections for D\O's $W+{\rm jets}$ and $Z+{\rm jets}$
are: one or two 
isolated high $p_T$ leptons with large missing $E_T$ (for $W$), and two jets
with $p_T > 20$ GeV and $|\eta| < 2.5$.
Figure~\ref{fig:wz} a) and b) show the $\Delta R_{\rm jj}$ 
and dijet mass distributions
from the D\O\ $W$+jets sample, based on a intergrated luminosity 
of 35 pb$^{-1}$. 
The spectra are normalized to the same total yield. There
is good shape agreement between data and Monte Carlo.
We are working to understand the uncertainties of the
theoretical prediction of the total yield.
Figure~\ref{fig:wz}c) shows
jet multiplicity distribution from the CDF $Z$ + jets sample, based on 
a intergrated luminosity of 57 pb$^{-1}$, for jet $E_T > 15$ GeV 
and $|\eta| < 2$, 
with and without
$b$-tagging (secondary vertex tag).  There is a good 
agreement between observed and expected number of $b$-tagged jets.

\subsection{$H\to WW^*$ searches}
In the SM
$H\to WW^*$ is the dominant decay channel if the Higgs mass is greater 
than 135 GeV. 
The cross section times branching ratio is largest for Higgs mass around 
160 GeV. A fourth generation fermion family would enhace the Higgs cross 
section by about factor of 8.5 relative to the SM for a Higgs boson 
mass range of 100-200 GeV~\cite{fourgen}.
In fermiophobic/Topcolor Higgs models the coupling to fermions is suppressed,
so these models give a larger branching ratio to boson 
pairs~\cite{topcolor}. 
The cleanest diboson decay channels are those where both $W$'s decay
leptonically. 

The main backgrounds for diboson Higgs searches are 
$Z/\gamma^*$, $WW$, $t\bar{t}$, $W/Z$+jets, and QCD multijet productions. 
Requiring large missing tranverse energy can remove much
of the
$Z/\gamma^*$ background. Another useful discriminant variable
is the opening angle between the two charged leptons, 
$\Delta\phi_{\ell \ell}$.
The two charged leptons from Higgs decay tend to move parallel due
to spin correlations in $H\to WW^*$ decay products 
thus these $H\to WW$ events have smaller $\Delta\phi_{\ell \ell}$ 
than background events.

\begin{table}[t]
\caption{Data and Expected background as a function of cuts in the D\O\
$H\to WW^*\to e\nu e\nu$ analysis. The cuts are optimized for $M_H = 120$ GeV.
\label{wwtable}}
\vspace{0.4cm}
\begin{center}
\begin{tabular}{|c|c|c|}
\hline
Event Selection &
Expected Background &
Data
\\ \hline
Lepton ID, $p_T>10, 20$ GeV &
$2748\pm42\pm245$ &
2753
\\ \hline
$M_{ee} < M_H / 2$ &
$264\pm18.6\pm4.3$ &
262 
\\ \hline
missing $E_T > 20$ GeV &
$12.3\pm2.5\pm0.7$ &
11 
\\ \hline
$M_T > M_H + 20$ GeV &
$3.6\pm1.4\pm0.2$ &
1 
\\ \hline
$\Delta\phi_{ee} < 2.0$ &
$0.7\pm1.4\pm0.1$ &
0
\\ \hline

\end{tabular}
\end{center}
\end{table}

\begin{figure}
\centerline{
\psfig{figure=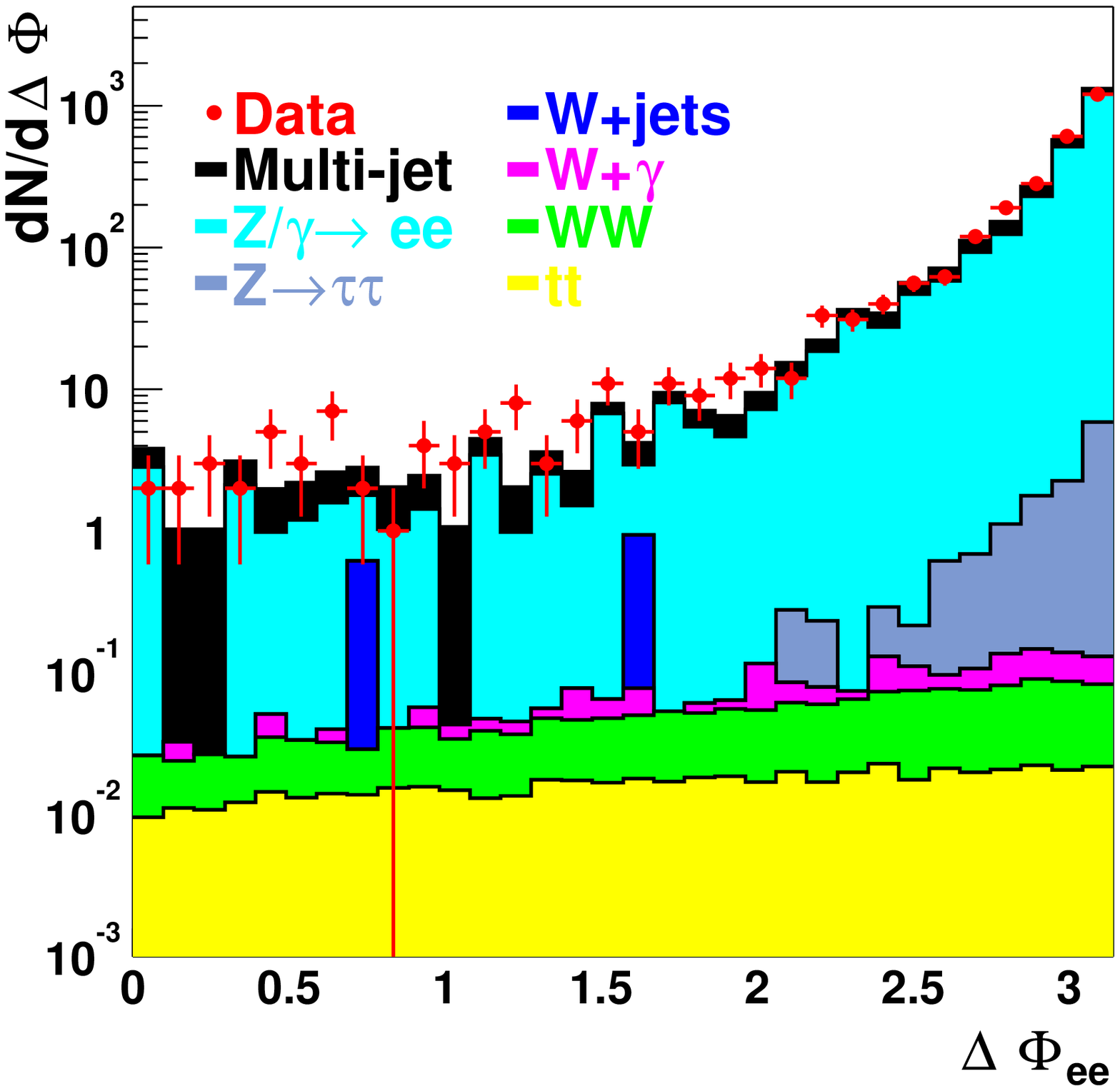,height=2.0in}
\psfig{figure=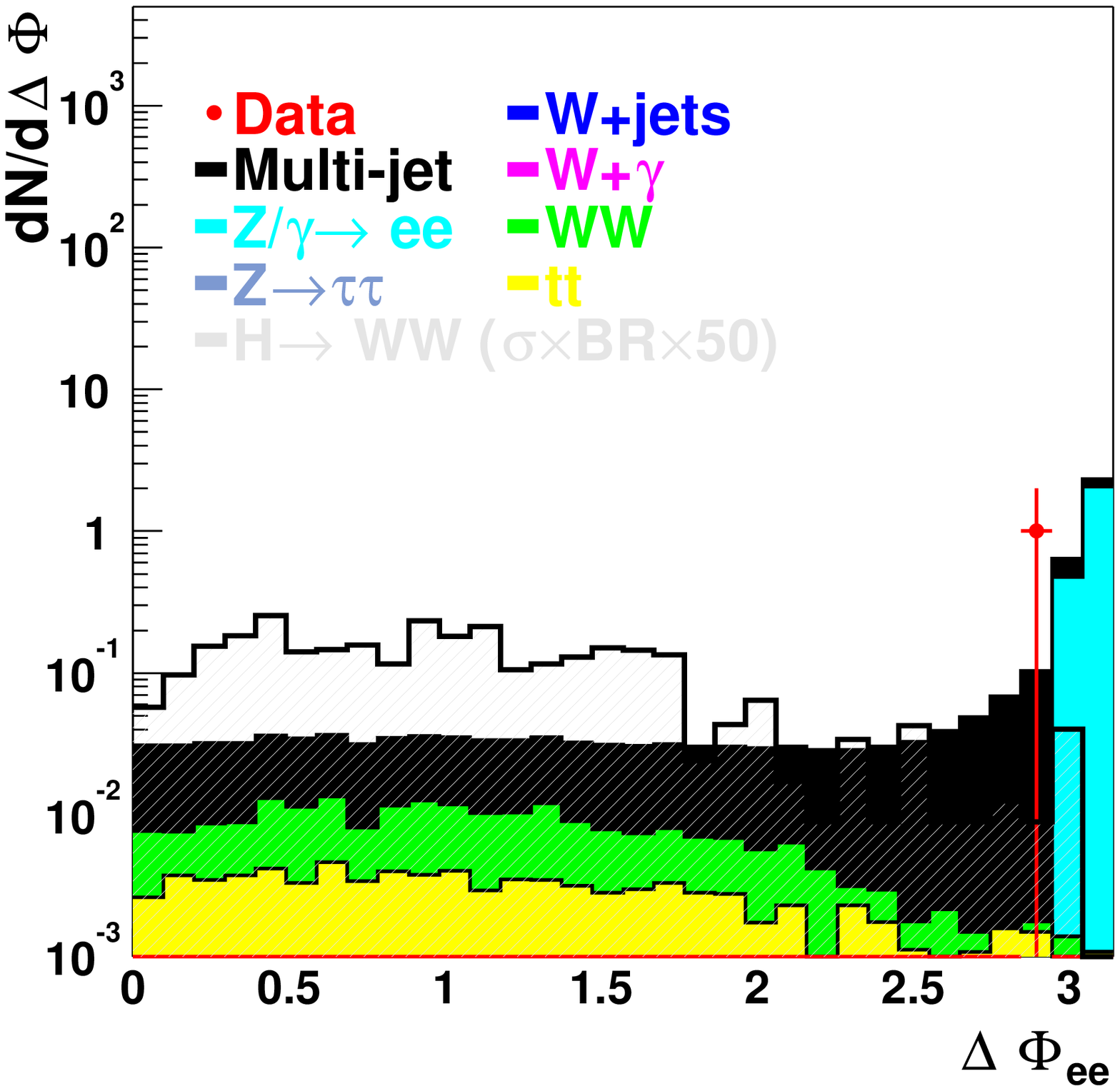,height=2.0in}
\psfig{figure=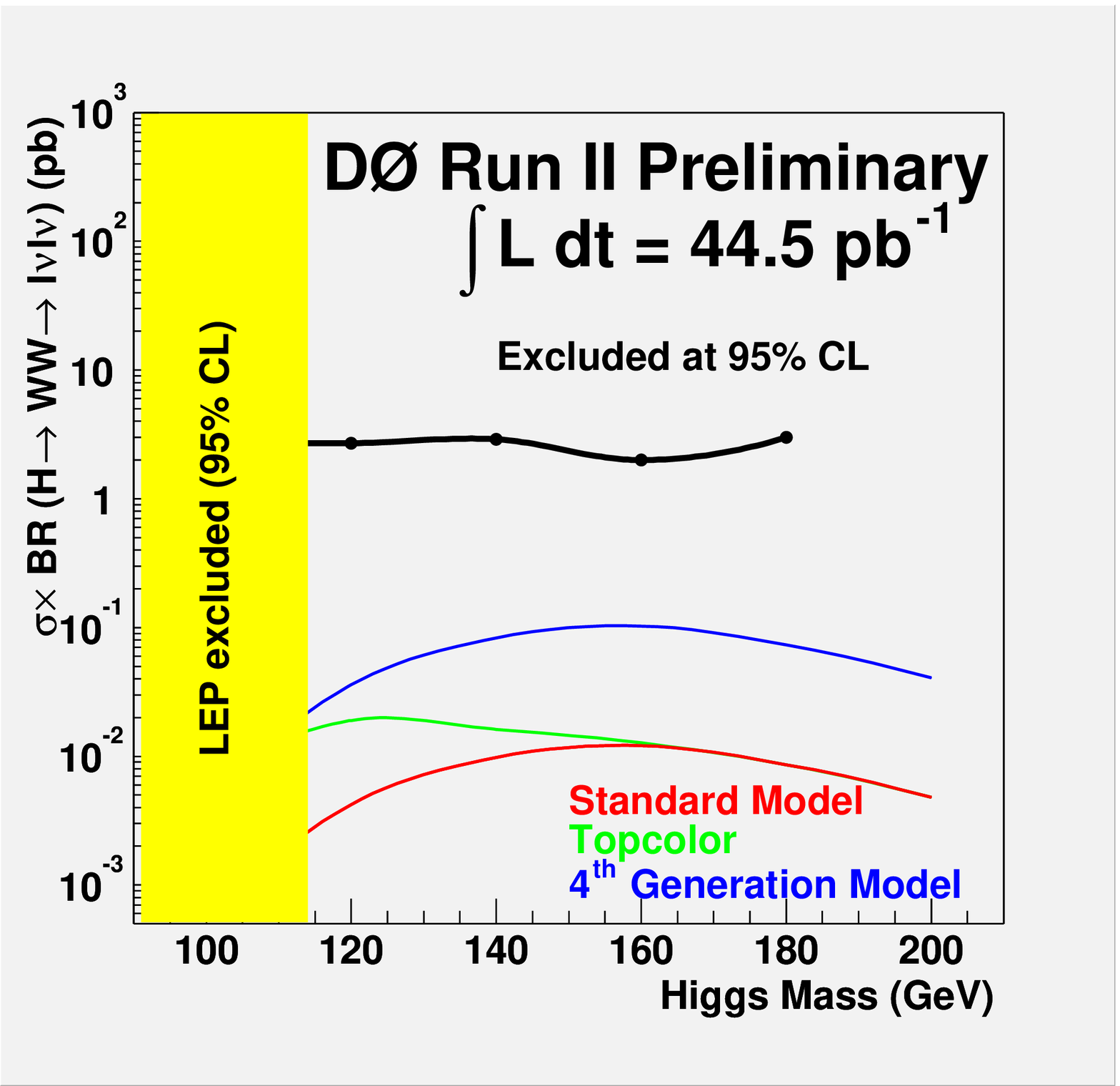,height=2.0in}}
\vskip -0.1cm
\centerline{a) {\hskip 5cm} b) {\hskip 5cm} c)}
\vskip -0.1cm

\caption{a) and b) are $\Delta\phi_{ee}$ distributions for data and 
backgrounds for $H \to WW^* \to e\nu e\nu$ searches, 
a) is after basic kinematic cuts and b) is after selection cuts.
c) is the cross section times branching ratio of $H \to WW \to l\nu l\nu$
versus Higgs mass.The black line is the D\O\ upper limit at $95\%$ C.L.; 
the color lines indicate expectations from various models.
\label{fig:hww}}
\end{figure}

The data used for D\O\ analysis of
$H\to WW^*\to e\nu e\nu$ search was collected from
Sept. 2002 to Jan. 2003, corresponding to an intergrated luminosity of 44.5 
pb$^{-1}$. Table~\ref{wwtable} shows the event selection optimized for
$M_H = 120$ GeV. The efficiency for this selection is about 8\%. 
Figure~\ref{fig:hww} a) shows the $\Delta\phi_{ee}$ distribution for
data and background processes after basic kinematic cuts, which mainly
are lepton identification and lepton $E_T$ requirements. 
The data and background shown in Fig.~\ref{fig:hww} a) are in agreement in
rate and shape.
Figure~\ref{fig:hww} b) 
shows the $\Delta\phi_{ee}$ distributions  
after all selection cuts listed in Table~\ref{wwtable} except the 
$\Delta\phi_{ee}$ cut. There is only one event from data in this plot,
which is removed by the 
$\Delta\phi_{ee}$ cut.
Fig.\ref{fig:hww}c) shows the $\sigma(p\bar{p}\to H) \times Br(H\to WW)$
limit at 95\% C.L. from current data, along with the expectations from 
the standard model, Topcolor, and fourth generation model.

\subsection{Searches for Double Charged Higgs}
Doubly-charged Higgs bosons appear in exotic Higgs representations 
such as found in left-right symmetric models \cite{hpp}. At the 
Tevatron the doubly-charged Higgs
can be produced in pairs through $\ppbar\to Z\gamma X \to H^{++}H^{--}X$, or
produced singly through WW fusion.
CDF searched for doubly-charged Higgs by looking 
at leptonic decays of $H^{++}$,
using same-sign electron pairs. 
The search region is above 100 GeV.
The search is performed in an invariant mass window of $\pm 10$\% 
of a given $H^{++}$ mass, which is about
$3\sigma$ of the detector resolution. 
The main backgrounds are from $Z$'s, QCD, and $W$+jets.

\begin{figure} 
\centerline{
\psfig{figure=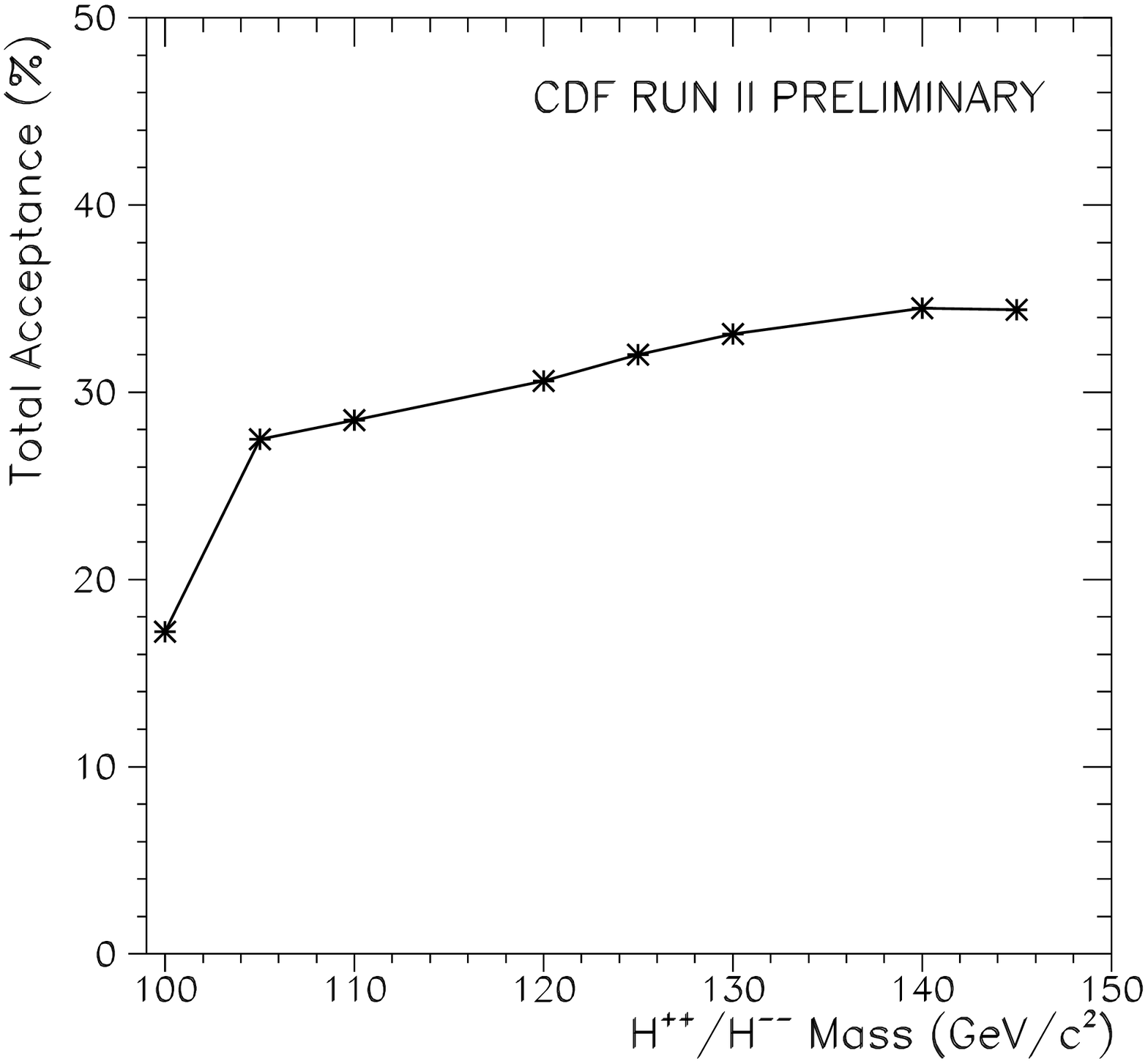,height=1.8in}
\psfig{figure=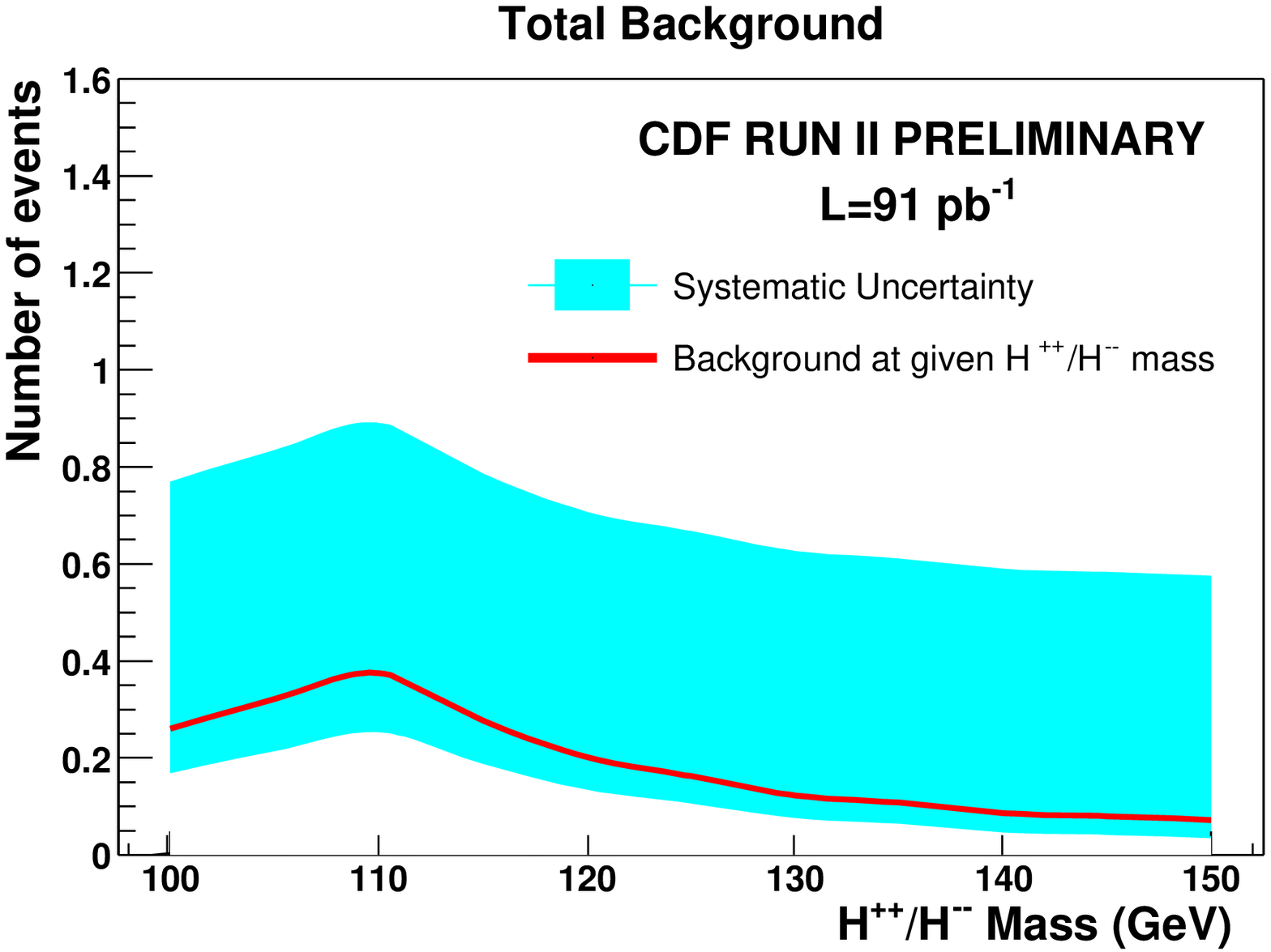,height=1.8in}
\psfig{figure=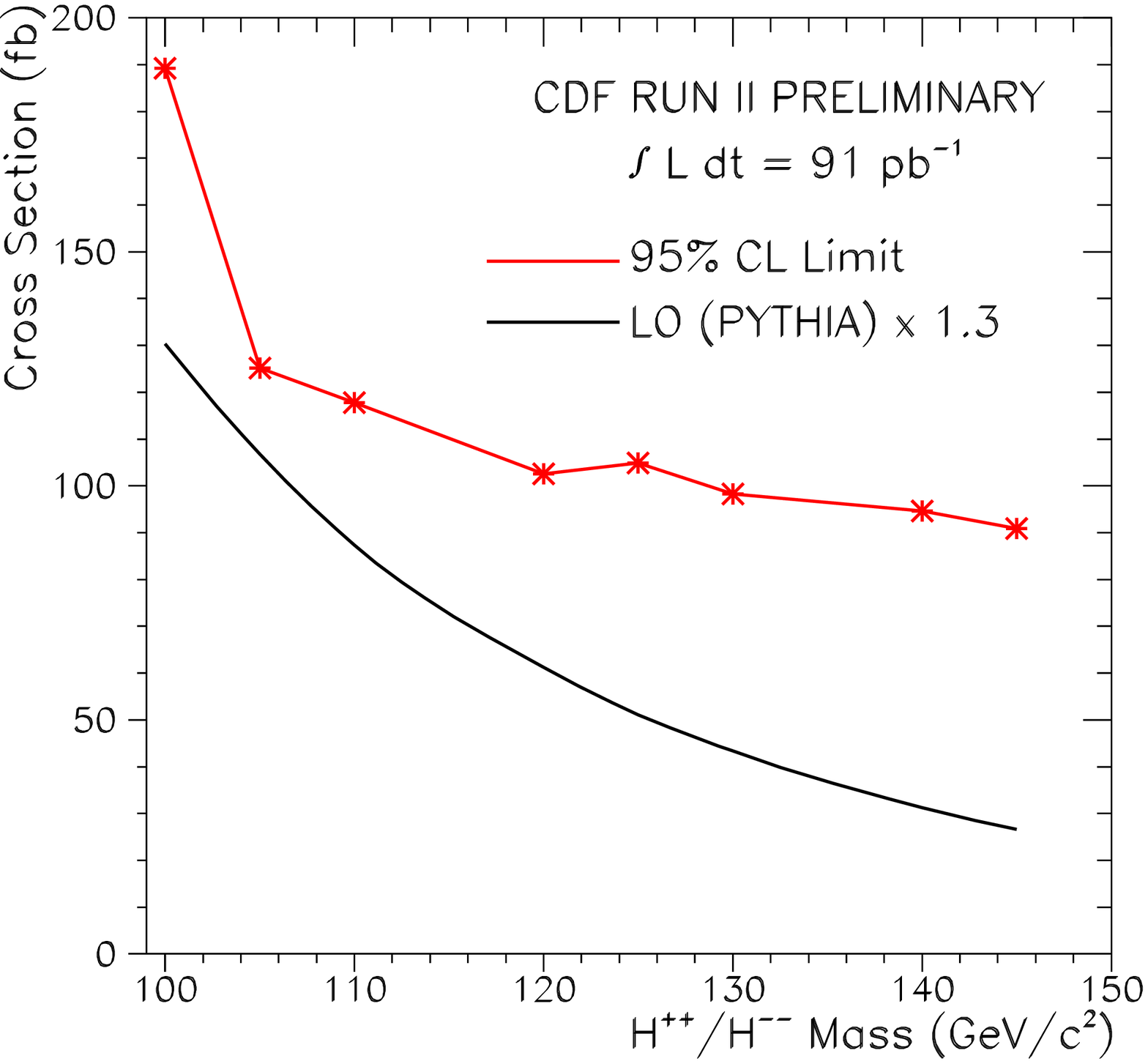,height=1.8in}}
\vskip -0.1cm
\centerline{a) {\hskip 5cm} b) {\hskip 5cm} c)}
\vskip -0.1cm
\caption{
a) The CDF total acceptance for doubly-charged Higgs, using the Pythia event 
generator and a GEANT-based detetctor simulation,
b) the total background as function of the Higgs mass, and
c) the production cross section and 95\% C.L. upper limit.
\label{fig:hpp}}
\end{figure}

In the 80-100 GeV mass range the instrumental background from $Z$ production
is dominant. The background occurs when one of the electrons from $Z$ 
radiates a photon, which subsequently converts.  When the
wrong sign conversion track is associated with the electron cluster, the 
event is reconstructed with two same-sign electrons.
The $Z$ background is estimated using the data in the 80-100 GeV mass 
range and the search covers the region above 100 GeV.  
The mass region of 100--130
GeV is dominated by the high-mass tail of the $Z$; above 130 GeV, QCD and 
$Z$ production processes 
are expected to contribute equally to the background.
Figure~\ref{fig:hpp} a) and b) show the acceptance and background 
for doubly-charged Higgs as a function of Higgs mass.
The large upper band in Fig.~\ref{fig:hpp} b) for the systematic uncertainty 
arises from the $W$+jets
background, for which CDF predicts 0.0 +0.5 $-$0.0 events.

The low mass region ($<80$ GeV) is used as a test of the background prediction.
CDF predicts 0.6 events in this region and observes 0 events.  In the search 
region ($>100$ GeV), CDF also observes 0 events. 
The Fig.\ref{fig:hpp} c) shows the 95\% C.L. cross section upper limit 
for pair-production of doubly-charged Higgs, based on an intergrated
luminosity of 91 pb$^{-1}$.

%

\section*{Acknowledgments}
I thank the Moriond organizers for an exciting and enjoyable conference.
I thank my colleagues from the CDF and D\O\ collaborations for sharing
their analyses results.

\end{document}